**Unique Spectroscopy and Imaging of Mars with JWST**

Villanueva[*] (NASA/CUA), Altieri (IAPS/INAF), Clancy (SSI), Encrenaz (LESIA), Fouchet (LEISA), Hartogh (MPS), Lellouch (LEISA) López-Valverde (IAA), Mumma (NASA), Novak (Iona), Smith (NASA), Vandaele (BISA), Wolff (SSI), Ferruit (ESA), Milam (NASA)

**Abstract**

In this document, we summarize the main capabilities of the James Webb Space Telescope (JWST) for performing observations of Mars. The distinctive vantage point of JWST at the Sun-Earth Lagrange point (L2) will allow sampling the full observable disk, permitting the study of short-term phenomena, diurnal processes (across the East-West axis) and latitudinal processes between the hemispheres (including seasonal effects) with excellent spatial resolutions (0.07 arcsec at 2 µm). Spectroscopic observations will be achievable in the 0.7-5 µm spectral region with NIRSpec at a maximum resolving power of 2700, and with 8000 in the 1-1.25 µm range. Imaging will be attainable with NIRCam at 4.3 µm and with two narrow filters near 2 µm, while the nightside will be accessible with several filters in the 0.5 to 2 µm. Such a powerful suite of instruments will be a major asset for the exploration and characterization of Mars. Some science cases include the mapping of the water D/H ratio, investigations of the Martian mesosphere via the characterization of the non-LTE $CO_2$ emission at 4.3 µm, studies of chemical transport via observations of the $O_2$ nightglow at 1.27 µm, high cadence mapping of the variability dust and water ice clouds, and sensitive searches for trace species and hydrated features on the Martian surface. In-flight characterization of the instruments may allow for additional science opportunities.

* To whom correspondence should be addressed.
E-mail: geronimo.villanueva@nasa.gov

**Index**

## 1. Introduction

Orbiting and landed missions to Mars have revolutionized our understanding of the history and evolution of the planet. Recent observations reveal rapid changes of the atmosphere (e.g., Mumma et al. 2009; Smith 2004) and on the surface of Mars (e.g., Malin et al. 2006) that are particularly challenging to detect with current orbiting assets due to the limited cadence and repeatability of the sampled regions. The rich instrumental suite of the James Webb Space Telescope (JWST) will permit sampling Mars from an unprecedented perspective (Figure 1) at the second Sun-Earth Lagrange point (L2). Such orbital configuration allows sampling the full observable disk, ultimately enabling the study of short-term phenomena, diurnal processes (across the East-West axis) and processes between the hemispheres including seasonal effects.

Mars and Earth share many similarities, with both terrestrial planets having moderate atmospheric temperatures, a rich geological history and evidence of similar beginnings in which habitable conditions persisted in both planets. On the other hand, the Martian atmosphere/surface is currently an arid and highly inhospitable environment, as revealed by recent in-situ measurements (Grotzinger 2013) and astronomical observations (Villanueva et al. 2013). The upper layers of Martian soil have been heavily eroded by long-term eolian activity, and the lack of magnetospheric and atmospheric shielding has led to modified soil chemistry (especially organics) through exposure to energetic particles.

But there is ample evidence that ancient Mars was wet and likely hosted habitable conditions (Bibring et al. 2006; Carr 1999), particularly during the Noachian age (3.6-4.5 billion years ago), and probably leading to the formation of rich subsurface aqueous reservoirs. Moreover, the presence of extensive volcanism probably gave rise to widespread hydrothermal activity and the formation of diverse chemical environments. Subsurface hydrates surviving from that time could incorporate organics produced by geological processes or by living organisms. Measurements of epithermal neutron fluxes, obtained by Mars Odyssey (Feldman et al. 2004), suggest the presence of important near-surface hydrogen concentrations on Mars. This hydrogen is often interpreted as buried water ice in low-latitude sites; however, independent evidence is required to establish its chemical form (e.g., hydrocarbons and/or mineral hydrates). Within the Gale Crater, soils analysis carried by the CheMin (Leshin et al. 2013) and ChemCam (Meslin et al. 2013) aboard the Curiosity Rover found evidence that hydrogen was either adsorbed $H_2O$ or corresponded to the hydration of an amorphous low-$SiO_2$ component of the soils. More globally, a detection of enhanced hydrocarbons at these sites could test whether sub-surface hydrogen is chemically bound in hydrocarbon moieties, and would strengthen the possibility of a habitable environment in the sub-surface of Mars.

The right balance of energy sources, temperature, pressure and chemical diversity leads to prosperous environments for life on Earth. Thanks to the plentitude of recent discoveries of extremophile organisms, the limits for such conditions have greatly expanded. On Earth, cold-loving and cold-tolerant microorganisms have been discovered and in some cases resuscitated from cores taken from permafrost regions, sub-permafrost and ice deposits that range in age from 40 Kyr to several Myr (Nienow et al. 1988; Gilichinsky et al. 2003; Tung et al. 2005). Some of the microorganisms subsist on a simple set of chemical reactions and evidence suggests that they are actively producing gas at temperatures below the freezing point of water (Tung et al. 2005). Experiments performed on other isolates recovered from these deposits demonstrate that they are metabolically active at temperatures down to -15 to -20° C (Christner 2002). Mars may have regions in its subsurface permafrost that could harbor endolith communities.

In order to investigate the conditions of early Mars and the habitability of its sub-surface, sensitive isotopic (see Section 3.1) and comprehensive chemical surveys (see Section 3.4) provide unique windows into the past and into the protected habitable regions of the planet. For instance, the latest measurements of atmospheric D/H in water on Mars (Villanueva et al. 2008; Novak et al. 2011; Villanueva et al. 2015) reveal strong geographical and localized variability and a D/H much higher than that in Earth's oceans, indicative of a significant loss of water owing to the preferential escape of the lighter isotopic fraction over geologic times. Chemical imbalances can also indicate current activity and a possible connection with stored sub-surface reservoirs of organics and water. Therefore, comprehensive and sensitive surveys of trace species, as recently done (Villanueva et al. 2013), are key for setting constraints on the level of activity on Mars, and the stability of its atmosphere over time. Above all, because the different organic species have substantially different resident lifetimes in the Martian atmosphere (from hours to centuries), such measurements not only test for current release but also provide stringent limits on quiescent levels (see Section 3.4).

With its capability for hemispheric snapshots (maps) of chemical species and processes, JWST will also permit unprecedented investigations of atmospheric composition and processes on Mars. It can probe the mesospheric $CO_2$ non-LTE emission at 4.3 µm (see Section 3.3) but also measure $O_2$ nightglow (at 1.27 µm) with unique high temporal and spatial resolutions. By doing so, JWST will allow us to better understand the processes acting on the chemical stability of the Martian atmosphere (see Section 3.2). In the study of aerosols and of the Martian surface composition (see Section 3.5), JWST can map and monitor the variability dust and water ice clouds, and can search for hydrated features and for carbonates on the Martian surface. Importantly, all these studies will be performed with outstanding sensitivities from the L2 vantage point, allowing the creation of maps with high spatial resolution and short time cadences (see Section 2).

## 2. JWST Capabilities for observing Mars

### *2.1 Windows of observation and achievable spatial resolutions*

The James Webb Space Telescope (JWST) is a powerful infrared observatory, with a broad range of spectroscopic and imaging capabilities targeting the 0.6 – 28.5 μm spectral region. The primary mirror, with a diameter of 6.5 meters (25 m$^2$ collecting area), is composed of 18 hexagonal beryllium mirror segments, which are constantly adjusted to optimize focus and optical performance by an adaptive wavefront sensing and control process. The telescope will be placed at the second Earth-Sun Lagrange point (L2) at a distance of 1.5 million km (0.01 AU) from Earth pointed opposite to the Sun with a Field-Of-Regard (FOR) limit of 85° to 135° (see Figure 2). A complex and extensive Sun-shield composed of 5 layers of reflective material will prevent light from Sun/Earth/Moon from reaching the telescope and instruments and passively cool the observatory to provide an especially stable cryogenic environment.

The observatory will be launched in October 2018 onboard an Ariane 5 vehicle, and is expected to acquire "first light" 28 days after launch, with early science observations starting ~37 days after launch. Considering the restrictions of FOR, Mars observations are only possible every two years (approximately), a few months after/before Mars-Earth opposition, with the first possible opportunity ending on December 14th 2018 (see Figure 2). Subsequent observations are possible in mid-2020/early-2021, and late-2022/early-2023. A broad range of Mars seasons (Ls) will be sampled in the first three apparitions, ranging from Ls:9° (Northern Spring) to Ls:206° (Northern Summer) and Ls:319° (Northern Winter). Most windows will sample the Southern cap, and the polar caps are only observable when illuminated (due to geometry). Due to the FOR restriction, most of the observable disk will be in daytime (84-93%). The evening terminator will be observable when Mars is approaching, and the morning terminator when Mars is receding.

As shown in Figure 3, Mars will reach a maximum rate of non-sidereal motion (100 arcsec/hour) within the observable windows. The guiding and tracking system of JWST has been designed to accommodate non-sidereal rates (up to a maximum of 108 arcsec/hour), with a pointing accuracy of 0.005 arcsec and a stability 0f 0.02 arcsec (Norwood et al. 2014). Mars' angular diameter will change substantially during the observing windows, from 8 to 20 arcsec. JWST presents a resolution of 0.07 arcsec at 2 μm wavelength, so the mean spatial resolution there will range from 23 to 58 km at the center of the disk. Such spatial resolutions will allow resolving fast and localized phenomena (e.g. over craters, ancient volcanoes, basins), which combined with the capability to map the complete hemisphere will allow unique investigations of dynamical and transport processes acting on the Martian atmosphere.

## *2.2 Mars expected brightness and available observing modes*

Mars' intense brightness is a principal factor restricting its access to JWST. The main goal of the observatory is the study of distant and faint objects, so a neutral density filter to the optical path was excluded to prevent the possibility that it could remain permanently locked in the system. Mars brightness varies substantially across the planet, reaching up to 200 Jy/arcsec[2] at short-wavelengths (see Appendix 1) in regions of high-reflectivity (see Figure 4). The 0.6 to 4 μm spectral range is dominated by reflected sunlight, which varies across the planet due to surface albedo and solar airmass (behaving primarily as a Lambertian surface). Marginal fluxes are expected from the night-side regions of Mars at these wavelengths (see Figure 5).

At longer wavelengths ($\lambda$>4 μm), Mars brightness is mainly defined by the surface temperature, which ranges from 140K to 300K with a typical value of 210K. Strong $CO_2$ opacity and dust/ice extinction/scattering reduces the planet brightness in the 2.5 to 5 μm spectral region, favoring observations with JWST at these wavelengths. Observations with the Mid-Infrared instrument (MIRI, 5-28.5 μm) and the Near-Infrared Imager and Slitless Spectrograph (NIRISS) are not likely to be feasible for Mars due to saturation limits determined from pre-flight tests. Observations with the Near-Infrared Camera (NIRCam) are possible when using the narrow filters (see section 2.4) and with the Near-Infrared Spectrograph (NIRSpec) when employing the high-dispersion mode and sub-array readouts (see section 2.3).

The detectors have latencies of 0.02% 10 seconds after a saturating source is removed. The decay is relatively fast so long as the source only saturated the detector by a factor of a 2-3. At higher saturation levels other non-linear effects start to become active, and the decay of the latent image is much prolonged. As shown in Figure 4 for NIRSpec, saturation levels are expected to reach a maximum of a factor ~3 for the least favorable wavelengths, while NIRCam could reach considerable saturation when using the broad filters. In general, careful planning will be required when organizing Mars observations with JWST in order to avoid flux saturation across the observable disk.

## *2.3 NIRSpec spectroscopic capabilities*

The Near-Infrared Spectrometer (NIRSpec) is a grating spectrograph sampling the 0.7 to 5.2 μm spectral region with a maximum resolving power $\lambda/\delta\lambda$ of 2700, up to ~ 8000 in the 1.0-1.25 μm range when employing the high-resolution spectral mode. Full spectral coverage is achieved with three gratings spanning from 0.7 to 1.8 μm, 1.7 to 3.0 μm and 2.9 to 5.2 μm. NIRSpec provides a broad range of optical entrance slits and observing modes, including an Integral-Field-Unit (IFU) with a Field-Of-View (FOV) of 3 x 3 arcsec$^2$ (achievable with 30 image slices of 0.1 x 3 arcsec$^2$), a flexible Micro-Shutter-Array (MSA) with 4x175x365 (255,500) openings of 0.2x0.45 arcsec$^2$ each, and a set of fixed slits (0.4x3.8, 0.2x3.3 and 1.6x1.6 arcsec$^2$). For most wavelengths, saturation limits using the IFU mode cannot accommodate Mars' intense flux, and this mode is thus not recommended for Mars observations. The MSA system is not favored for extended and bright sources, since small leaks across the shutters would quickly add and overwhelm the detectors, and therefore observations of Mars employing this mode are not possible.

Observations of Mars are only feasible when employing the narrowest fixed-slit (0.2 x 3.3 arcsec$^2$) at the maximum resolution (2700 or 8000), and when performing sub-array/window readouts. The focal-plane array of NIRSpec is comprised of two 2048x2048 HgCdTe arrays (arranged along the dispersion axis) with a spatial sampling rate of 0.1 arcsec per pixel, and a corresponding broad FOV of 3.4 arcmin across the spatial axis. In Figure 4, the saturation limits for NIRSpec are presented for a sub-array readout of 2048 x 64 pixels using the narrowest slit at a resolving power of 2700. In general, NIRSpec observations are feasible with sub-array readouts in the 2.7 to 5.2 μm spectral region for most of the observable disk, and from 0.7 to 5.2 μm in the night regions (see Figure 5).

Mapping of Mars can be achieved by letting the planet shift under the slit at a constant rate. With typical non-sidereal motion rate of 60 arcsec/hour and a Mars diameter of 10 arcsec the telescope can crisscross Mars in 10 minutes, and by employing 3 staggered slit positions (3 x 3.3 = 9.9 arcsec) full disk coverage is obtained in a total of 30 minutes. Since Mars quickly fills the detector wells (fractional second), faster mapping is certainly possible.

The Near-Infrared Camera (NIRCam) is a dual-band imaging camera sampling the 0.6 to 5.0 μm spectral region (short: 0.6-2.3 μm, long: 2.4-5.0 μm) with a large range of narrow and broad spectroscopic filters. The focal-plane array consists of four HgCdTe 2048x2048 detectors for the short band, and one 2048x2048 detector for the long wavelength band. The system has been designed to be fully redundant, so a total of 10 detectors comprises the focal-plane array. The plate scale is 0.032 arcsec/pixel for the short-band and 0.065 arcsec/pixel for the long-band, with both bands having an extensive FOV of 2.2 x 2.2 arcmin$^2$. Mars observations with NIRCam are feasible with two narrow filters near 2 μm for the complete disk, and with several filters (from 0.5 to 2 μm) across the night regions (see Figure 4). Mapping of $CO_2$ non-LTE emission at 4.3 μm will be also possible with NIRCam's F430M filter.

## 3. Prime Mars Science Goals with JWST

Such a powerful suite of instruments will be a major asset for the exploration and characterization of Mars. In this section we summarize some science cases, yet in-flight characterization of the instruments may allow for additional science opportunities.

### 3.1 The evolution of water on Mars (D/H isotopic mapping)

Water is a key ingredient in the definition of habitability of a terrestrial planet. By learning about the evolution and the stability of the water reservoirs on Mars, we can better understand where and when habitable conditions persisted on the planet. Measurements of epithermal neutron fluxes obtained by Mars Odyssey (Feldman et al. 2004) suggest the presence of important near-surface hydrogen concentrations on Mars, while the main water reservoir on Mars, the Polar

Layered Deposits (PLD), is estimated to contain ~21 m global equivalent layer (GEL) of water (Zuber et al. 1998; Plaut et al. 2007; Lasue et al. 2013). The atmosphere holds a fraction of the labile water on the planet, yet it acts as a buffer between the exosphere (at the boundary with outer space) and the main reservoirs of H, C and O (e.g., regolith and polar caps), with atmospheric isotopic/abundance ratios providing key diagnostics for quantifying the exchange among these environments.

Isotopic ratios are among the most valuable indicators for the loss of volatiles from an atmosphere. Deuterium fractionation also reveals information about the cycle of water on the planet and informs us of its stability on short- and long-term scales. The vapor pressures of HDO and $H_2O$ differ substantially near the freezing point, making the condensation/sublimation cycle of the isotopologues sensitive to local temperatures and saturation levels and to the presence of aerosol condensation nuclei (Fouchet & Lellouch 2000; Bertaux & Montmessin 2001). Including this vapor pressure isotopic effect in Global Climate Model also suggests that the D/H ratio should diminish as water vapor is transported from sources to sinks (Montmessin et al. 2005). Local isotopic ratios such as derived by Curiosity/MSL (Webster et al. 2013) are representative of that specific location and time, and may not represent the actual D/H of atmospheric water on Mars (see more below), even though reports of hemispherically averaged D/H in water indicated a similar value of 5-6 times the D/H of Earth's oceans (VSMOW) – Owen et al. 1988; Bjoraker et al. 1989; Krasnopolsky et al. 1997. Importantly, these isolated (in time and in space) measurements of D/H in the atmosphere were typically – but incorrectly – assumed to be representative of the bulk atmosphere. Spatially resolved (~100 km) measurements of D/H at different times of day and seasons are necessary to disentangle local from global phenomena, in particular when studying transient related processes (e.g. cloud formation) over localized regions (e.g., craters, ancient volcanoes, basins). Such maps reveal the true isotopic ratio of current water reservoirs with implications for the global loss of water over geologic time, and may also assist in the identification of new sources of water on Mars.

Within the spectral region accessible with NIRSpec, both isotopologues of water, HDO and $H_2O$, have strong rotational-vibrational bands. Specifically, by targeting the $\nu_1$ fundamental band of HDO at 3.7 μm and several bands of $H_2O$ in the 2.5 μm to 3.4 μm spectral region, one can obtain sensitive and accurate retrievals of water D/H. Such bands have been targeted from ground-based observatories, permitting the global mapping of the D/H on Mars (Villanueva et al. 2008; 2015), although with limited spatial resolution (>0.7 arcsec). Furthermore, the spectral resolution provided by NIRSpec is sufficient to separate the molecular features (see Figure 6), while the broad spectral coverage permits improved correction for scattering and extinction by dust and ice clouds.

In Figure 6, we present synthetic spectra of HDO and $H_2O$ assuming a water abundance of 150 ppmv and 5 VSMOW, leading to strong absorptions (2% for HDO, 3-5% for $H_2O$). Considering that we expect photometric signal-to-noise-ratio in the thousands at these wavelengths with NIRSpec (http://jwstetc.stsci.edu/etc/input/nirspec/spectroscopic/), we would derive highly accurate D/H retrievals for this case; yet as recently demonstrated by Villanueva et al. 2015, the D/H varies substantially across the planet (1-10 VSMOW) while the water abundance by two orders of magnitude across the planet and along seasons (~5-500 ppmv). Nevertheless, the high photometric precision of JWST would allow us to test the D/H of water on Mars at very low water abundances (>10 ppmv, comparable to Villanueva et al. 2015), permitting to sample a substantial portion of the observable disk during most seasons at high-spatial resolution.

JWST could also be a great asset to monitor the main Martian water cycle. This cycle has been extensively characterized by the TES instrument on board MGS (Smith 2004) and shown to be highly reproducible from one Martian Year to another. However, the TES measurements were all obtained at a 2pm local time. As a result, the diurnal variability of the water vapor in the Martian atmosphere is still poorly constrained. By measuring the water vapor over a same location at dawn, noon and dusk, JWST will sample the Martian diurnal cycle. Moreover such observations could be coordinated with the ExoMars Trace Gas Orbiter whose orbit is designed to uniformly sample both the spatial and temporal evolution of Mars atmosphere.

Exceptionally, an interannual variability of the Martian water cycle has been observed, clearly associated with the onset and development of a global dust storm (Fouchet et al. 2011). JWST could be very valuable to search for any interannual variability of the water cycle and to identify their cause.

*3.2 Chemical processes and meridional transport*

Ozone and water are important species for understanding the stability and evolution of Mars' atmosphere. They are closely linked through photochemistry and they are anticorrelated with each other. This relationship was first observed by Mariner 9 (Barth et al. 1973; Conrath et al. 1973). Ground-based observers (Traub et al. 1979) confirmed this anti-correlation, and later measurements by the Viking orbiter revealed additional aspects of their variability (Jakosky & Farmer 1982). Clancy & Nair 1996 further examined the photochemical cycle and proposed that the anti-correlation between these two species also depends on altitude and used the seasonal variation of the "hygropause" (an altitude at which water vapor reaches saturation) to explain the observed seasonal cycles.

Photolysis of $O_3$ by UV sunlight produces $O_2(a^1\Delta_g)$ that has a lifetime of 1.25 hours (Ball et al. 1993). Detection of $O_2(a^1\Delta_g)$ emissions indicate the presence of ozone above that altitude (Novak et al. 2002; Altieri et al. 2009). This approach was first used to study ozone on Earth (Noxon 1968) and was then extended to Mars (Noxon et al. 1976).

The near-infrared oxygen band emission at 1.27 μm occurs in all three terrestrial atmospheres, and is particularly prominent in the Mars atmosphere, where two distinct processes lead to vertically deep, globally extended $O_2(a^1\Delta_g)$ emission. Figure 7b presents the latitudinal/vertical cross-section of Mars $O_2(a^1\Delta_g)$ volume emission rates as simulated by the LMD (Laboratoire de Météorologie Dynamique) GCM (Global Climate Model) photochemistry (Lefèvre et al. 2004). Like Earth, Mars sunlit latitudes (northward of 60S in Figure 7b) present $O_2(a^1\Delta_g)$ dayglow emission associated with the solar photolysis of ozone, which peaks in cold atmospheric regions due to its anti-correlation with ozone-destroying atomic hydrogen, a photolysis product of water vapor (e.g., Barth et al. 1973). Consequently, Mars $O_2(a^1\Delta_g)$ dayglow enables critical studies of Mars oxygen and hydrogen photochemistry, as well as atmospheric water vapor content/saturation conditions. Like Venus, the upper atmospheric $O_2(a^1\Delta_g)$ nightglow on Mars (southward of 60S above 40 km in Figure 7b) results from upper level global transport of atomic oxygen, which recombines to form $O_2(a^1\Delta_g)$. As such, Mars polar winter $O_2(a^1\Delta_g)$ nightglow provides a unique tracer of cross-hemispheric meridional circulation in the Mars upper atmosphere.

Mars $O_2(a^1\Delta_g)$ nightglow has only recently been observed from spacecraft limb observations (Clancy et al. 2013; Bertaux et al. 2012), whereas Mars $O_2(a^1\Delta_g)$ dayglow has been monitored from both ground-based (e.g., Novak et al. 2002; Krasnopolsky 2003) and spacecraft platforms (Fedorova et al. 2006). JWST has the potential to observe new aspects of both Mars $O_2(a^1\Delta_g)$ dayglow and nightglow in support of Mars photochemical and dynamical studies.

While JWST saturation limitations preclude direct observation of Mars $O_2(a^1\Delta_g)$ dayside emission, JWST angular and spectral resolutions would provide the first opportunity to measure the nightside (evening) falloff of $O_2(a^1\Delta_g)$ dayglow emission associated with the 1.25 hour lifetime for $O_2(a^1\Delta_g)$ radiative decay (see Figure 7a). $O_2(a^1\Delta_g)$ is also subject to collisional de-excitation at altitudes below ~20 km, a reaction which is poorly constrained by available measurements. Specifically, within the observable JWST windows (Figure 2), we can sample 2 to 3.4 hours of nighttime (7-16% of the observable disk) beyond the day/nigh terminator. Hence, unique JWST observations of the evening decay of $O_2(a^1\Delta_g)$ dayglow support photochemical and collisional reaction rate determinations. JWST sensitivity and angular resolution also support unique spatial mapping of $O_2(a^1\Delta_g)$ polar nightglow in nadir (non-limb) geometry (Figure 7a). Such observations of polar $O_2(a^1\Delta_g)$ nightglow would contribute unique studies of spatial and temporal variability associated with meridional transport into Mars high latitudes during winter.

### 3.3 Upper atmosphere sounding (non-LTE $CO_2$ and mesospheric $CO_2$ ice clouds)

The numerous ro-vibrational bands of $CO_2$ at 4.3 μm produce a strong absorption feature, clearly seen in all available observations of the three terrestrial planets. It is also well known that during daytime there is an emission feature in the center of this spectral region due to solar fluorescence in the upper atmosphere (Lellouch et al. 2000; Alvarado et al. 2013; Formisano et al. 2005a). López-Valverde et al. 2005 carried out an analysis of this emission feature on Mars, as observed by ISO and PFS/Mars Express, using a state-of-the-art non-Local Thermodynamic Equilibrium (non-LTE) model. They showed that the major contribution arises between 80-120 km above the surface (Figure 8a) and is due to a large number of ro-vibrational bands of $CO_2$ in this spectral region. A major contribution comes from the second hot bands ($v_3 + 2v_2 \rightarrow 2v_2$), excited by direct solar pumping at 2.7 μm and partially relaxed via emission at 4.3 μm. This produces the double lobed characteristic emission (see Figure 8). The emission varies considerably with airmass, being very small for low solar illumination. The fraction of collisional quenching and radiative relaxation involving so many isotopic and high energy states of $CO_2$ is subject to some uncertainty, since the relaxation rates are not all well determined.

The non-LTE emission can be considered as a limitation for sounding the atmospheric temperature from the $CO_2$ absorption band. As an example, nadir observations at 4.3 μm by NIMS/Galileo and by VIRTIS/Venus Express have been used in the past to derive the Venus thermal structure but only during nighttime (Roos-Serote et al. 1995; Grassi et al. 2010). However, this emission feature offers the possibility to sound the upper atmosphere (e.g., Sonnabend et al. 2012), a poorly known region on Mars.

We consider JWST to be an excellent tool to study this emission in order to further our understanding of the fluorescent excitation of $CO_2$ and subsequent cascading. Furthermore, such studies will permit to derive information about key atmospheric parameters in the Martian upper mesosphere, such as high-altitude temperatures and their global variability. NIRSpec at its higher spectral resolution will permit sounding of the upper regions of Mars' atmosphere at these wavelengths, through mosaics of the whole planet attained at the largest spatial resolution achievable, so that a global high-resolution picture of the Mars mesosphere is obtained for the first time.

To test the possibility to study the mesospheric temperature of Mars, we consider here a first sensitivity study of the nadir-looking emission to be observed by NIRSpec to changes in the Martian mesosphere using the non-LTE model mentioned above. An arbitrary change of 40 K was introduced on a given thermal structure between 50 and 100 km. Maximum solar illumination (sub-solar point) and an average emission angle of 60 degree were assumed. Figure 8b shows the calculated spectra for the NIRSpec highest resolution (R~2700). Due to the non-LTE nature, the local temperature has only an indirect and relatively small effect on the IR emission, mostly by changing the line shapes and the initial solar pumping. Rate coefficients also vary with temperature and this also has an additional and also small effect. The final result is an increase of 0.006 W m$^{-2}$ sr$^{-1}$ um$^{-1}$ (0.017 Jy for a 0.2" slitwidth and 0.1" pixel pitch, see conversions in Appendix 1), or 20-25 %, at the emission peak.

The high sensitivities provided by JWST will enable high signal-to-noise ratios at these wavelengths with NIRSpec (~0.0001 Jy sensitivity in 1 minute), allowing measurements of thermal structure in the upper mesosphere of Mars with errors around 5 K (only limited by modeling uncertainties, see below) – unprecedented for a global observation of the upper atmosphere of Mars. Two results are expected, the actual absolute temperatures at each pixel and the spatial variability of those temperatures in an instantaneous snapshot of the whole planet. This second objective may be more robust, as it may be less impact by potential biases, and will test our current understanding of the mesosphere as a whole.

In addition to the obvious difficulties that arise from the actual noise, pointing stability and sensitivity eventually achieved by NIRSpec, there are other difficulties that need to be handled in order to achieve the goals mentioned above. They arise mainly from geometrical considerations and from the uncertainties in diverse parameters, which will remain undetermined during the observations. We list here some of them and propose approximations and tentative solutions which will require a more detailed study. These difficulties may produce biases in the absolute temperatures at the level of individual pixels, but will have smaller impact on the atmospheric variability to be studied at a planetary scale.

The simulations in Figure 8b used very different and extreme assumptions for the surface solar reflection, from a strong reflection to a pure thermal component. They produced very different results outside the range 2270-2390 cm$^{-1}$ (4.4-4.2 $\mu$m), where the atmosphere is optically thinner, but the results in the center of the band are robust and independent of the surface conditions or dust scattering. This is consistent with the result that the non-LTE emission originates high in the atmosphere. Therefore, in principle this should not be an important source of uncertainty.

A traditional difficulty for mesospheric sounding using a nadir geometry is the unknown state of the troposphere if this has a significant contribution. The 4.3 $\mu$m daytime observations do not permit derivation of the tropospheric temperature. The contribution to the total non-LTE emission may not be entirely negligible. Therefore, a priori information from climatologies like the Mars Climate Database will be needed. The uncertainty associated with those climatologies is much larger at high altitudes.

López-Valverde et al. 2011 showed that with good spectral resolution the non-LTE emission at 10 $\mu$m, ultimately arising from solar pumping mechanisms similar to the 4.3 $\mu$m emission, may be up to 10 times larger in the limb than in nadir. The actual factor is smaller if the spectral resolution is poorer (Gilli et al. 2009; Formisano et al. 2006). In principle this limb enhancement is good for sounding purposes if the limb could be sampled, but unfortunately the minimum pixel size of NIRSpec is not sufficient to separate nadir and limb contributions. The exploitation of the nadir observations will therefore require a correction for the limb contribution. This can be a numerically expensive exercise but can be handled carefully and properly by current non-LTE models. The pixels with large contributions from the limb may not be useful even with the proposed correction, although they will contain useful information for specific studies like non-LTE model validation.

Within the core of the strong $CO_2$ band at 4.3 μm, OMEGA detected the $CO_2$ ice band slightly offset with respect to the gaseous band. This reflection is due to mesospheric $CO_2$ clouds. A climatology has been established for these clouds using OMEGA and HRSC dataset (Määttänen et al. 2010) and their occurrence has been shown to be significantly influenced by the wave activity in the mesosphere (Spiga et al. 2012; Listowski et al. 2014). The global coverage and the spectral resolution larger than OMEGA spectra provided by JWST, will help to further characterize the global distribution of the CO2 mesospheric clouds, their particle sizes and shapes.

*3.4 Organics*

The first sensitive survey for volatile organic compounds ($CH_4$, $C_2H_2$, etc.) and other trace species was obtained with the moderate resolution ($\lambda/\Delta\lambda \sim 100\text{-}1000$) Mariner 9 infrared spectrometer (5-50 μm), measuring upper-limits for thirteen trace species (Maguire 1977). By the 1980s, advances in ground-based astronomy permitted sensitive searches for certain sulfur species (e.g., Encrenaz et al. 1991) and other trace gases, including hydrogen peroxide ($H_2O_2$, Clancy et al. 2004; Encrenaz et al. 2004; 2012; Hartogh et al. 2010) and methane (e.g., Krasnopolsky et al. 1997). Yet the prime region to search for hydrocarbons and several other species is the 2-5 μm infrared region, where the C-H vibrational stretch and other fundamental vibrational modes reside for a plethora of trace species (composed of some combination of elements C, H, N, O, P, and S) – see Villanueva et al. 2013. Only one instrument had probed this spectral region from Mars orbit (PFS/MEx, Formisano et al. 2005b), although the instrument suffers from organic contamination and complex ringing (introduced by micro-vibrations) in the C-H spectral region (~3 μm). In situ, the SAM instrument on the Curiosity Rover recently detected chlorobenzene (150–300 ppbw) and $C_2$ to $C_4$ dichloroalkanes (up to 70 ppbw) whose sources could possibly be atmospheric, geological, biological or external (Freissinet et al. 2015).

Detection limits improved significantly with the advent of extremely sensitive high-resolution infrared array spectrometers at large ground-based telescopes and advanced data analysis techniques, ultimately permitting the detection of methane ($CH_4$) in 2003 (Mumma et al. 2009, ppbv levels). The recent observations of methane by four groups (Mumma et al. 2009; Formisano et al. 2004; Krasnopolsky et al. 2004; Fonti & Marzo 2010) indicate regions of localized release, and high temporal variability. Sensitive upper-limits for $CH_4$ and several other trace species (of potential biological and geological origin) were derived from a multi-year ground-based campaign (Villanueva et al. 2013). More recently, in-situ measurements obtained by the Curiosity rover also reveal strong variability of methane on Mars, perhaps indicative of nearby releases (Webster et al. 2015, average $CH_4$~0.7 ppbv abundances with peaks of 12 ppbv).

The existence of volatile organics on Mars would indicate recent release, in particular, when considering the reported variability. Such claims are hard to reconcile with the static view once assumed for Mars, and this combined with the complexity of the observations have prompted questions on the reliability of the detections (Zahnle et al. 2011; Zahnle 2015). Are organics being released into the Martian atmosphere? Such question requires accurate spectroscopy, high-spatial resolution and global coverage in order to identify regions of possible active release.

NIRSpec/JWST has comparable spectral resolution to PFS/MEx, yet improved sensitivity, reduced systematic effects and instantaneous hemispheric global coverage would permit sensitive searches for organics and other trace species in the 3-5 μm. Specifically, JWST can obtain parts-per-billion (ppbv) limits on organics ($CH_4$, $CH_3OH$, $H_2CO$, $C_2H_6$, $C_2H_2$, $C_2H_4$), hydroperoxyl ($HO_2$), nitrogen compounds ($N_2O$, $NH_3$, HCN) and chlorine species (HCl, $CH_3Cl$) by targeting their fundamental bands in the 2.7-3.3 μm spectral region (see Figure 9 for a selected set of species and additional details in Villanueva et al. 2013).

*3.5 Variability of surface composition and aerosols (dust and water clouds)*

In recent years our knowledge of the Martian surface composition and geomorphology has strongly improved thanks to the scientific return of experiments on board space missions such as Mars Global Surveyor (MGS), Mars Odyssey, Mars Express (MEx) and Mars Reconnaissance Orbiter (MRO). Remote sensing data with increasing spatial resolution have allowed the identification of those minerals crucial to infer the role of water on the surface of the planet. The details achieved have made it possible to associate the corresponding geomorphological features with an unprecedented precision.

On a global scale, mafic materials usually match with darker units. TES/MGS (Christensen et al. 2001) and OMEGA/MEx (Bibring et al. 2004) instruments mapped (ortho- and clino-) pyroxene-rich rocks on the ancient cratered terrains of the southern hemisphere and in the Syrtis Major area (e.g., Bandfield et al. 2003; Poulet et al. 2005; Mustard et al. 2008; Carrozzo et al. 2012). Olivine has been clearly detected preferentially in the Nili Fossae, Syrtis Major and Hellas regions, as well as around several spread craters mostly located in the southern hemisphere (e.g., Hoefen et al. 2003; Mustard et al. 2005; Poulet et al. 2007; Carrozzo et al. 2012; Ody et al.

2012). Bright regions are characterized by a dusty soil rich in $Fe^{3+}$ nanophasic oxides, with no evidence of hydration features (e.g., Poulet et al. 2007). Hematite-rich terrains have been clearly observed in Sinus Meridiani, in Aram Chaos and smaller deposits inside Valles Marineris (e.g., Christensen et al. 2001; Glotch & Rogers 2007; Arvidson et al. 2006). The widest deposit of gypsum has been found around the Northern Polar Cap (Langevin et al. 2005), while smaller deposits have been mapped in Sinus Meridiani (e.g., Wiseman et al. 2008) and inside Valles Marineris (e.g., Gendrin et al. 2005; Kuzmin et al. 2009). Phyllosilicates have been detected in the Mawrth Vallis, Syrtis Major, Aram Chaos and other sparse areas in the exposed material around some craters in the Southern hemisphere (e.g., Poulet et al. 2005; 2007; Loizeau et al. 2010) and in the northern plains (Carter et al. 2010).

On a more local scale, the CRISM/MRO spectrometer confirmed those findings, showing a higher mineralogical diversity (Murchie et al. 2009), showing a higher mineralogical diversity. Other hydrated materials have been discovered on the surface, such as carbonates (e.g., Ehlmann et al. 2008) and opaline (e.g., Skok et al. 2010). OMEGA and CRISM data have demonstrated that at different spatial scales, the very ancient basaltic crust of Mars must have been altered by aqueous processes producing a diverse assemblage of hydrated phyllosilicate minerals. The combination on a global scale of the spectroscopic evidence of hydrated minerals with small-scale geomorphology features has revealed that the history of Mars is characterized by a rock cycle that can involve aqueous processes, burial and exhumation of materials.

In-situ measurements performed by the NASA Mars Exploration Rovers (MERs), Spirit and Opportunity, have provided further evidences that aqueous processes took place on Mars (Rice et al. 2010). Currently, even more constraining findings that habitable conditions formerly occurred on the planet have been collected by Curiosity Rover, operating on the Gale Crater site and travelling toward the layered deposits at the flanks of Mount Sharp. Data collected by Curiosity at Yellowknife Bay confirm that the sedimentary rocks contain basaltic minerals, Ca-sulfates, Fe oxide/hydroxides, Fe-sulfides and amorphous material (Vaniman et al. 2014). These rocks show a significant amount of clay minerals providing strong indications of an ancient fresh water environment (Grotzinger 2013).

JWST will be an extremely useful observatory for investigating Mars on a global scale, providing seasonal as well as short term evolutionary studies of surface albedo, airborne dust and water ice clouds. The Martian surface is variably masked by the presence of airborne dust and water ice clouds. A thin layer of dust is always present in the Martian atmosphere. Its amount can strongly change in the atmosphere, depending on the season and location on the planet, causing local, regional, and global dust storms (e.g., Cantor 2007). The long-term record of ground-based and orbiter observations indicates that the largest regional dust storms develop during the southern hemisphere spring and summer, when Mars is closer to the Sun (i.e., around perihelion). These events can grow and involve the entire planet, forming a global dust storm that is characterized by extremely high dust optical depths and lifting to high altitudes (above 50 km). However, global dust storms exhibit large interannual variability, occurring on average every third or fourth Martian year (e.g., Smith 2004). They can also show significant variability in their seasonal timing and amplitude, which are not well simulated by Mars Global Climate Models (MGCM). Furthermore, mechanisms that trigger the formation and evolution of both regional and global dust storms, such as surface lifting and redistribution of dust, require qualitative approaches that limit predictive model simulation.

In terms of remote sensing and radiative effects, airborne dust usually increases the reflectance of dark regions, while decreasing albedos of bright regions, and smoothing mafic and hydration diagnostic features in general. Dust particles suspended in the atmosphere strongly affect the Martian thermal structure and consequently influence the global atmospheric circulation (Haberle et al. 1982). Global and seasonal monitoring of the Martian albedo is crucial to better understand and constrain the role of dust in the Martian climatology. Accurate retrievals of dust opacity are only possible when the photometric behavior of the underlying surface is well known, although dust estimates could be derived by using the slope of the spectrum between 1-1.6 µm and 3-3.6 µm regions. Additional aerosol information (i.e., optical depth, the scale height, and optical properties) can be obtained by probing several distinct spectral features in the 2.6–3.0 µm spectral region as demonstrated by Fedorova et al. 2002.

In addition to dust circulation, Mars climate is also characterized by an active hydrological cycle, with large spatial and seasonal variations of water vapor, ice clouds, and surface frost. Every summer, water sublimes from the residual ($H_2O$) and seasonal ($CO_2$) polar ice caps, forcing large increases in global atmospheric water vapor that is in turn transported via global meridional circulation to be incorporated in the winter polar ice cap of the opposite hemisphere. Three main types of water ice clouds have been recognized on Mars (e.g., Clancy & Nair 1996; Tamppari et al. 2000; Pearl et al. 2001; Smith 2002; Smith et al. 2003; Smith 2004; Wang & Ingersoll 2002; Mateshvili et al. 2007; Benson et al. 2003): 1) the north and south polar hoods; 2) the aphelion cloud belt; 3) isolated orographic clouds above the major volcanoes. Polar water ice clouds form mainly near the edge of the seasonal polar cap in the corresponding spring and late summer/early fall. They are related to the deposition and sublimation of the seasonal polar caps. The aphelion cloud belt occurs from northern early spring through mid-summer and extends between (about) 10° and 30° N. It originates from the condensation of water vapor raised by the ascending branch of the Hadley circulation (Clancy & Nair 1996). Orographic water ice clouds have long been associated with volcanoes on Mars. However, only recent space missions could monitor their full seasonal and inter-annual behavior. Olympus, Ascraeus and Pavonis Mons present cloud activity between Ls = 0° and 220° with a peak around Ls = 100°. Water ice clouds over Alba Patera are mainly observed for Ls = 60° and Ls = 140°, with a minimum around Ls = 100°. Arsia Mons (the southern-most volcano) presents nearly continuous cloud activity throughout the Mars year (e.g., Benson et al. 2006).

Mars water ice clouds play important roles in Martian climate, influencing Martian atmospheric circulation (Madeleine et al. 2012; Wilson & Guzewich 2014) and the water vapor cycle (Montmessin et al. 2004). Moreover, the process of cloud formation plays an important role in the removal of atmospheric dust particles, which serve as condensation nuclei and subsequently precipitate together with water ice (e.g., Clancy & Nair 1996). The particle size of Mars water ice clouds plays a key role in all of the above processes. Water ice clouds show diagnostic features in the 3-3.6 μm region that can be employed to determine their particle sizes (e.g., Madeleine et al. 2012). The detailed temporal study of Mars clouds and particle sizes, together with the retrieval of dust opacity, is thus crucial to better understand and model the Martian climate and circulation (e.g., Wolff & Clancy 2003).

Time domain spectra of the 3-5 μm spectral range can reveal changes in surface hydration and in the surface temperature (e.g., Milliken et al. 2007; Audouard et al. 2014). The 3 μm surface hydration feature is observed everywhere on the planet. It has been thus interpreted as liquid water adsorbed or tightly bound in the structure of the materials of the near-top surface. Thanks to OMEGA data, a strong correlation has been also observed with clay outcrops and around the Northern polar cap. Spectral features at wavelengths > 3 μm in both TES (Bandfield et al. 2003) and PFS (Palomba et al. 2009) data have been interpreted as likely due to carbonates in the dust (though not yet definitive). However, OMEGA data did not confirm those findings, probably due to a different spectral resolution between OMEGA (20 nm) and PFS (1 nm). The identification of key minerals such us clays and carbonates, which formation is strictly connected to the presence of liquid water, has been possible in the 1-2.5 µm spectral range, as well as global maps of pyroxenes and oxides have been performed by means of spectral index retrieved in the 0.5-2.5 µm spectral range.

In the study of the Martian surface composition and its geomorphology, JWST can map and determine the temporal variability of the following parameters: dust opacity (1-1.6 µm & 3-3.6 µm); water ice clouds opacity (3-3.6 µm); 3 µm surface hydration band depth; surface temperature (4-4.6 µm); and the 3.4 µm band depth to detect carbonates in the airborne dust.

## 4. Synergy of JWST with current and planned planetary assets

### 4.1 Mars orbiters and rovers: MAVEN, MSL, ExoMars 2016, 2018, Mars 2020

JWST will expand and complement the capabilities of current assets at Mars (e.g, MAVEN, *Curiosity*) and also those planned in the near future (e.g., ExoMars 2016 & 2018, Mars 2020). For instance, measurements of D/H on Mars are among the main goals of the *Curiosity* Rover (MSL) and MAVEN missions, but both platforms provide local measurements only (near-surface in Gale Crater by MSL, and ionosphere by MAVEN) and do not provide a measure of the total atmospheric column across the planet. MAVEN has a sophisticated suite of instruments studying the processes responsible for the escape of species (e.g., H and D) from the Martian atmosphere, with measurements spanning a broad range of altitudes (150-6200 km). JWST observations will nicely complement these measurements by probing the integrated columns (0-100 km) of $H_2O$ and HDO, and by probing the thermal structure of the mesosphere (50-100 km) via $CO_2$ non-LTE emission. JWST, which is expected to provide measurements of Mars from 2018 to at least 2023 (Figure 2), will also complement and expand the observations by ExoMars/TGO (2016) beyond the mission lifetime (2016-2019).

The ExoMars mission includes several elements to be sent to Mars in two launches: The ExoMars Trace Gas Orbiter (TGO) with an Entry and Descent Demonstrator, called 'Schiaparelli' are planned for 2016. In 2018 a lander will deliver a rover to the Martian surface. The nominal Science Phase of TGO will end in November 2019, although the spacecraft will continue acting as a relay until end of 2022. Two suites of spectrometers will be on board TGO (NOMAD and ACS), performing a complete investigation of the composition and structure of the Martian atmosphere. One of the many objectives of TGO is to obtain maps of several trace gases of importance, such as $H_2O$ and its isotopologue HDO, or CO and methane and other constituents related to their chemistry. Vertical profiles of these trace gases as well as temperature will also be observed. Both NOMAD and ACS instruments are sensitive in the IR spectral region that will also be probed by JWST/NIRSpec, covering the 0.7 to 1.7 μm region with the ACS-NIR channel, and the 2.2-4.3 μm interval with the NOMAD-SO and LNO channels, some bands being also recorded by the ACS-MIR channel. The spectral resolving power of these instruments will exceed 10,000.

Coordinated observation campaign between the spectrometers on board ExoMars TGO and JWST would allow the cross-validation of the instruments, and it is especially important for tracking transient phenomena (e.g., methane outgassing). That is a unique advantage of JWST over TGO when building large-scale maps for the study of rapid phenomena, since global maps obtained by TGO will be built up over many orbits (requiring much more time than a JWST observation of the entire visible disc), and consequently what would appear as a spatial variation in an orbiter's map could be in fact be a time variation. Furthermore, JWST will permit building long-term time series beyond the lifetime of the ExoMars mission, key when trying to understand the complexity of the physical and chemical processes prevailing on Mars.

*4.2 Ground-based observatories: E-ELT, TMT, Keck, VLT, ALMA, IRTF*

Excellent synergy is expected between JWST and current (e.g., Keck, VLT, ALMA, IRTF) and future (e.g., E-ELT, TMT) ground-based observatories. JWST will provide superior spatial resolution and access to spectral regions that are opaque from the ground (e.g., non-LTE $CO_2$ emission), while ground-based observatories will provide improved spectral resolution. Even at the outstanding diffraction limits of E-ELT (0.02 arcsec at 3 μm), TMT (0.03 arcsec at 3 μm) and even Keck (0.07 arcsec at 3 μm), performing adaptive optics (AO) on an extended source like Mars is particularly challenging. With current AO technologies, the best corrections achieved spatial resolutions not better than 0.4 arcsec, as demonstrated with attempts to lock the adaptive optic systems of VLT and Keck on Mars (see Villanueva et al. 2010), yet future new developments in extreme AO may allow to overcome the current limitations. By combining the spatial resolution of JWST with the spectral resolution provided by powerful ground-based spectrometers, simultaneous observations will address the science goals described in section 3 with unprecedented accuracy and sensitivity.

JWST will also nicely complement observations acquired at radio wavelengths with ALMA. The large sub-millimeter array can provide spatial resolutions (in the 84 to 720 GHz spectral range) comparable to those achievable with JWST. Interferometric observations of an extended source requires a compromise between the spatial resolution and the spatial extent of the source in order to achieve proper calibration of the spectral images. At a resolution of 0.6 arcsec, continuum losses across the Martian disk with ALMA are small, yet improved resolutions (0.07 arcsec) would require detailed mapping and support by the ACA (Atacama Compact Array) to achieve proper calibration. Importantly, by inverting the lineshapes measured with ALMA one can characterize the vertical structure of the column densities measured with JWST. Such synergy between JWST and ALMA will be of great value, for instance, in the study of the D/H ratio on Mars, with JWST providing high-resolution spatial maps of the integrated columns of $H_2O$ and of HDO, and ALMA providing the vertical structure of these condensable species. Similar studies can be performed on CO lines (e.g., Villanueva 2004).

## 5. Conclusions

Upon its arrival to the L2 Lagrange point in 2018, JWST will open a new window on the exploration and characterization of Mars. JWST will provide unique sensitivities and excellent spatial resolution, while the orbital configuration will permit sampling the full observable disk, enabling the study of short-term phenomena, diurnal processes (across the East-West axis) and latitudinal processes between the hemispheres (including seasonal effects). JWST high sensitivities will actually pose challenges when observing certain wavelengths due to saturation constraints, yet a broad range of spectral ranges in the 1-5 μm region will be accessible with NIRSpec, including imaging with NIRCam. Many science cases can be attained with this powerful instrumental suite, and here we summarize some key investigations that would lead to extraordinary science: 1) high-resolution [space x time] mapping of the D/H ratio in Mars' water, enabling a sensitive search for unrecognized reservoirs of water on Mars and revealing the true deuterium enrichment in labile water (unaffected by climatological effects); 2) unprecedented investigations of the Martian mesosphere by probing the $CO_2$ non-LTE emission at 4.3 μm and $O_2$ nightglow at 1.27 μm, pertaining to radiation, photochemistry, and meridional transport in the Martian middle atmosphere; 3) mapping and monitoring the variability dust and water ice clouds, and searches for hydrated features and for carbonates on the Martian surface. Importantly, these studies will be synergistic and will enhance the science return from a broad range of current and future Mars assets (e.g. MAVEN, ExoMars) and ground-based observatories (e.g. ALMA, ELTs).

**Appendix 1: Flux conversions**

In this paper, the preferred unit of spectral flux is surface brightness [Jy arcsec$^{-2}$] because Mars is an extended source and fills the entrance slit of the spectrometers and the instrument pixels of the cameras. Importantly, the solar reflected component ($\lambda$<4 μm) of Mars brightness only varies across a Martian year due to Mars's orbital eccentricity by ±20% (when considering a constant albedo), while the thermal surface brightness can be considered constant when adopting a constant temperature. On the other hand, Mars spectral irradiance ([Jy], [Jansky], [$10^{-26}$ W m$^{-2}$ Hz$^{-1}$]) varies substantially (by orders of magnitude) according to the varying distance between Earth (JWST) and Mars. As shown in Figure 3, Mars' angular diameter varies typically from 8 to 20 arcsec within the observing windows of JWST.

Mars observations are only possible with NIRSpec when using the narrowest slit (0.2 arcsec), which would correspond to a solid angle of $\Omega$ = 0.02 arcsec$^2$ (4.7x10$^{-13}$ sr), when considering NIRSpec's focal-plane array pitch of 0.1 arcsec. Flux levels [Jy] reaching a NIRSpec pixel can therefore be computed from [Jy arcsec$^{-2}$] by multiplying by $\Omega$ [arcsec$^2$], while conversion to spectral radiance would be L [W m$^{-2}$ μm$^{-1}$ sr$^{-1}$] = 3x10$^{-12}$ E [Jy] / ( $\Omega$ [sr] $\lambda^2$ [μm$^2$]) = 0.13 B [Jy arcsec$^{-2}$] / $\lambda^2$ [μm$^2$].

Sensitivity limits for NIRSpec can be computed using the online exposure time calculator: http://jwstetc.stsci.edu/etc/input/nirspec/spectroscopic/

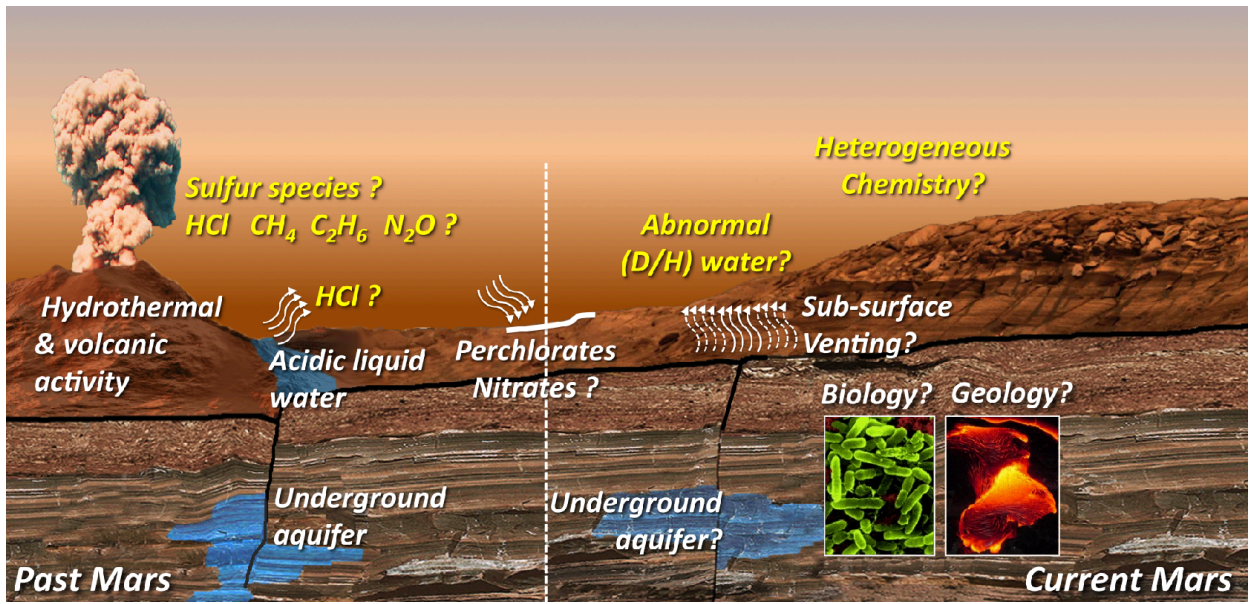

**Figure 1:** JWST will be able to address fundamental questions about Mars: How wet and habitable was Mars? How much water is currently available? How is the exchange between the polar caps with the regolith and the atmosphere? Are there sources of chemical disequilibrium (e.g. biology/geology)? How active is heterogeneous chemistry and what is the role of the engulfing dust storms? (Figure adapted from Villanueva et al. 2013).

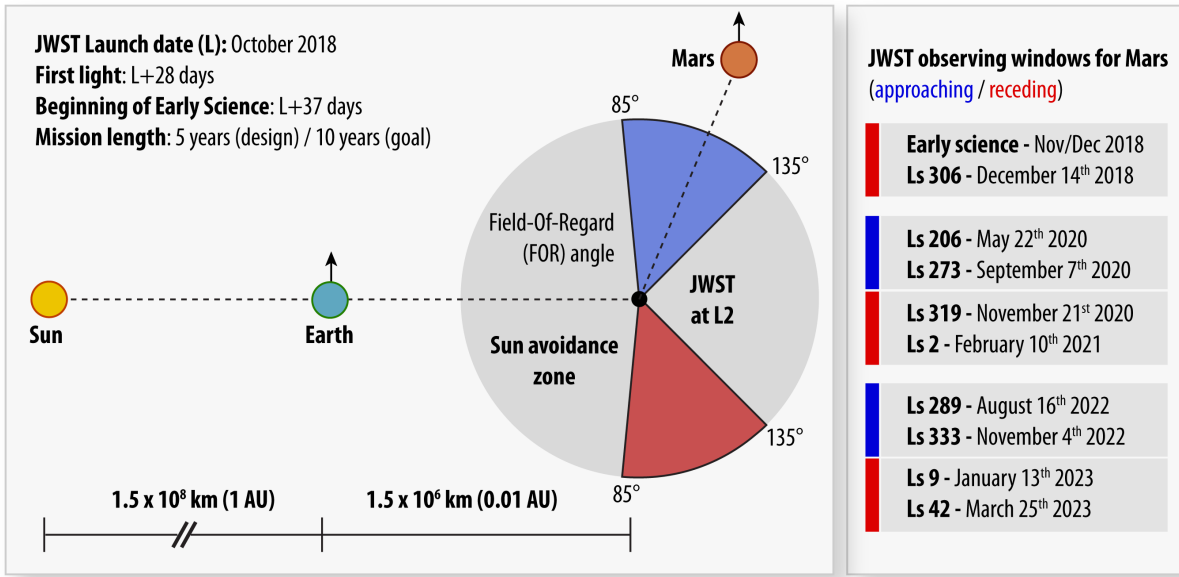

**Figure 2**: Observability of Mars with JWST. The observatory can perform observations within a Field-Of-Regards (FOR) of 85° and 135°, restricting the observations of Mars to months before/after the Mars-Earth opposition, occurring approximately every two years.

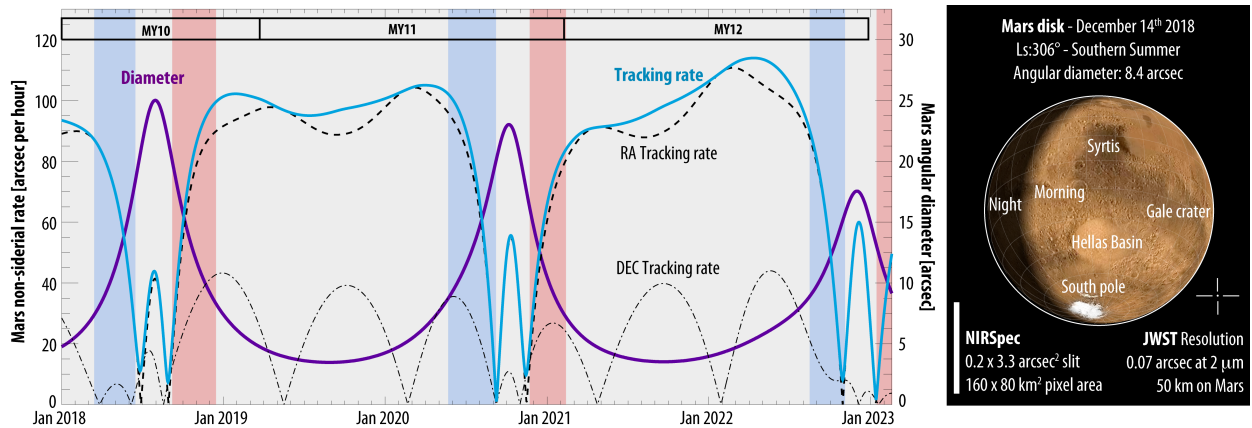

**Figure 3**: Evolution of Mars ephemeris during the first years of JWST. The purple trace indicates Mars angular diameter [right axis], while the light-blue trace indicates the total non-siderial rate of motion of Mars [left axis]. As explained in Figure 2, Mars is only observable in the blue/red windows. During these windows, Mars angular diameter will vary substantially (and herewith the achievable spatial resolution) from 8 to 20 arcsec. The maximum tracking capability of JWST is 108 arcsec/hour for non-siderial objects, a value within the rate of motion for Mars during all observable windows.

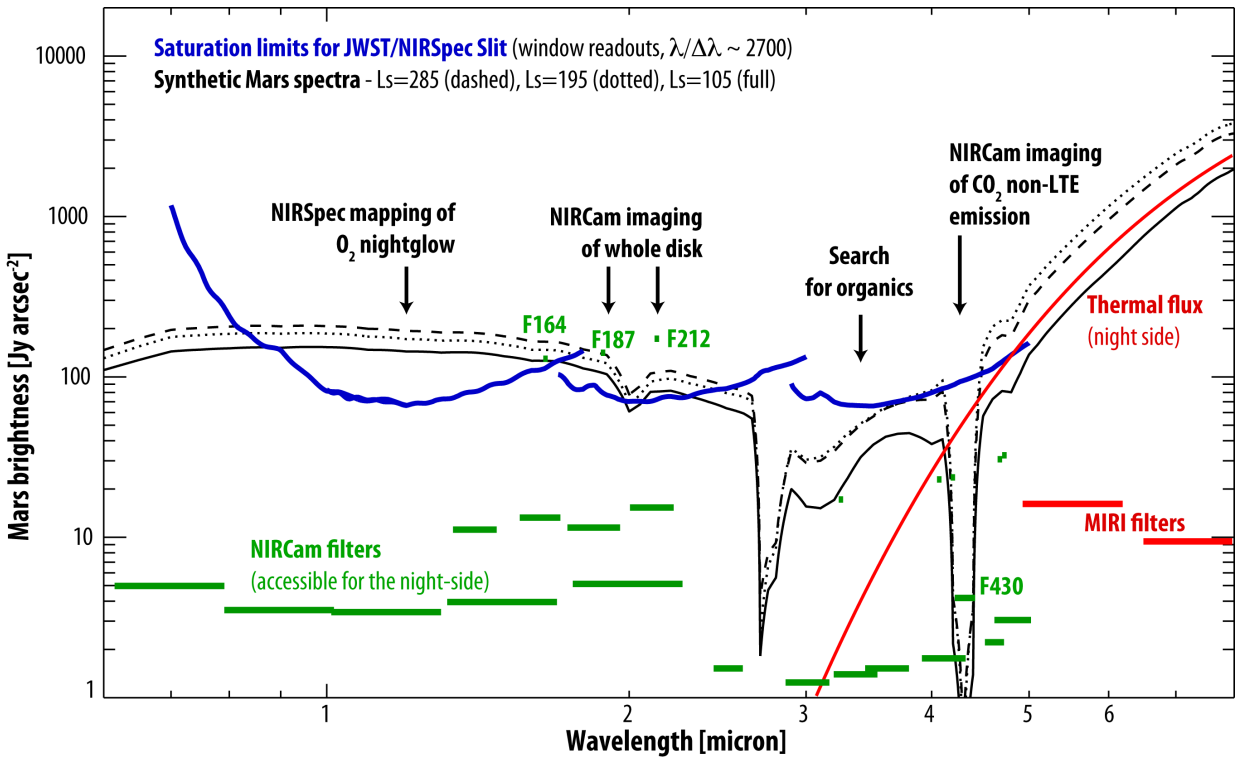

**Figure 4**: Predicted Mars brightness as seen from JWST for three seasons (Ls, aerocentric longitude), and the corresponding non-saturating NIRSpec limits. NIRSpec observations are feasible with sub-array readouts in the 2.7 to 5.2 μm spectral region for most of the observable disk (flux ranges are worst-case scenario), and from 0.7 to 5.2 μm in the night regions. Observations with NIRCam are accessible with two narrow filters near 2 μm for the complete disk, and with several filters from 0.5 to 2 μm across the night regions. Mapping of $CO_2$ non-LTE emission at 4.3 μm will be also possible with NIRCam's filter F430.

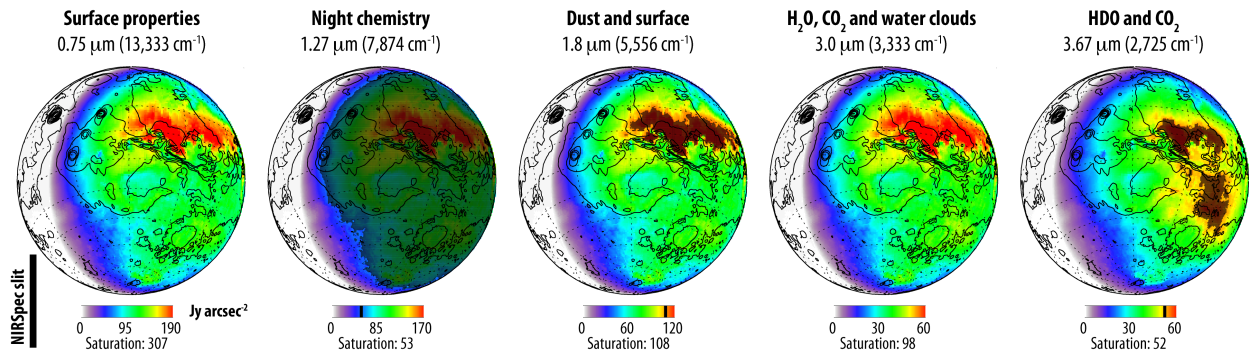

**Figure 5**: Simulated Mars continuum fluxes for five key spectral regions. The observing geometry and climatological conditions were computed for December 13th 2018 17:20 UT, when Mars is at a distance of 1.433 AU from the Sun and the season is Ls=306º (Southern Summer). All fluxes are in [Jy arcsec$^{-2}$], while the saturation levels are for spectroscopic observations with NIRSpec when considering sub-array readouts (see Figure 4). The night side is depicted as the white region, and in this instance the morning terminator is observed. Contours of the planet's topography (Smith et al. 2001) are overlaid to assist in the identification of Martian regions. NIRSpec's 0.2 x 3.3 arcsec$^2$ slit is shown for comparison, while additional spatial resolution and tracking information is presented in Figure 3.

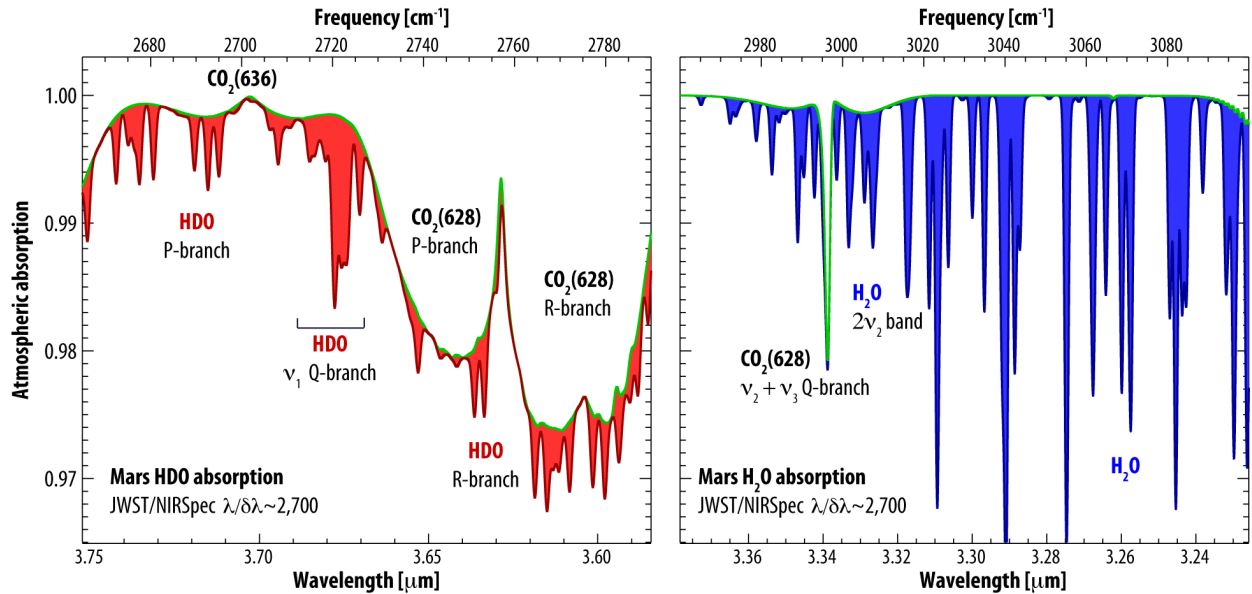

**Figure 6**: Simulated Mars atmospheric absorptions of HDO and $H_2O$ as observed with NIRSpec/JWST. The simulations consider a combined airmass (Sun-Surface+Surface-JWST) of 3 (1.5 Sun-Surface plus 1.5 surface-observer), a molecular abundance of 150 ppmv, a D/H of 5 VSMOW, and typical atmospheric pressures and temperatures for Mars. In both spectral windows, lines of isotopic $CO_2$ also allow probing the total atmospheric column, of critical value when computing atmospheric abundance ratios.

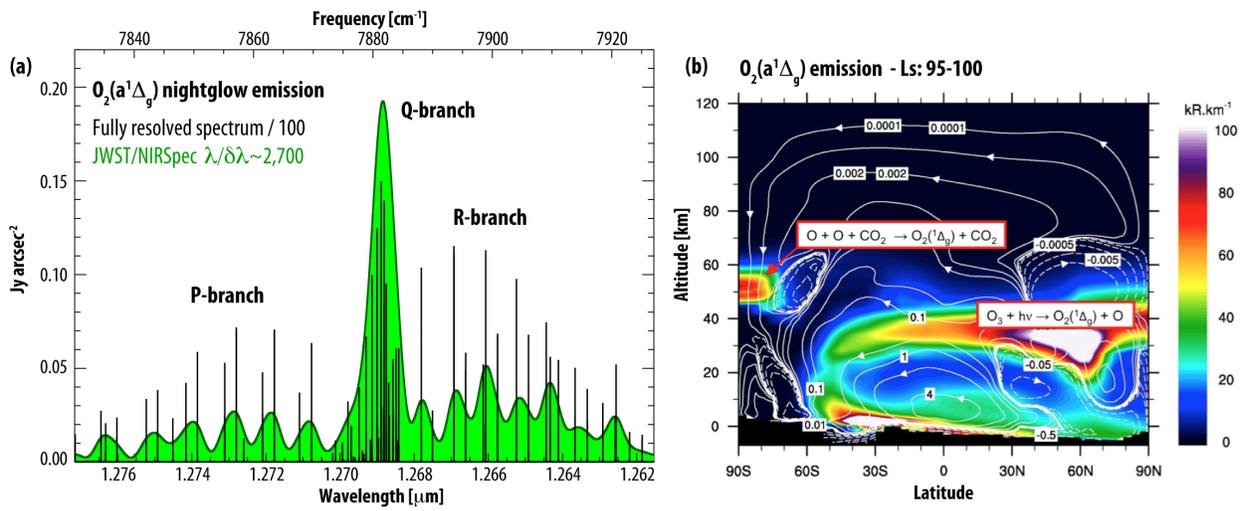

**Figure 7**: (a) Synthetic $O_2(a^1\Delta_g)$ nightglow emission on Mars convolved to the instrumental resolution of JWST/NIRSpec. A fully resolved spectrum is overlaid showing the numerous encompassing lines of $O_2$ emitting in this spectral region. (b) The vertical/latitudinal distribution of Mars $O_2(a^1\Delta_g)$ emission as simulated by the Mars LMD GCM photochemical model (Lefèvre et al. 2004). Two distinct processes lead to $O_2(a^1\Delta_g)$ production over winter polar (O recombination) and sunlit ($O_3$ photolysis) latitudinal regions (reproduced from Clancy et al. 2012).

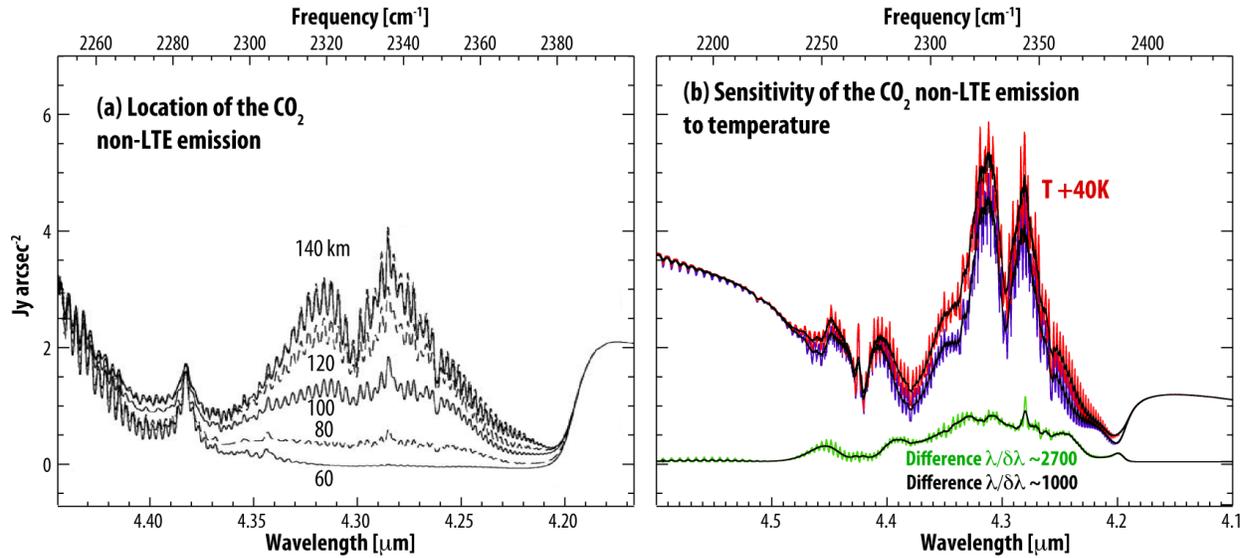

**Figure 8**: Characterization of non-LTE emission on Mars with JWST. (a) Simulation of the non-LTE emission as detected by a hypothetical observer located at different altitudes above the planet; the maximum contribution comes from the emission occurring above 80 km (after López-Valverde et al. 2005). (b) Sensitivity of the Martian non-LTE nadir emission at 4.3 μm to changes in the mesospheric temperature. Blue: nominal calculation for a given atmospheric state and with a strong solar scattering in the surface. Red: increase of 40 K in the mesospheric temperature and no scattering at the surface.

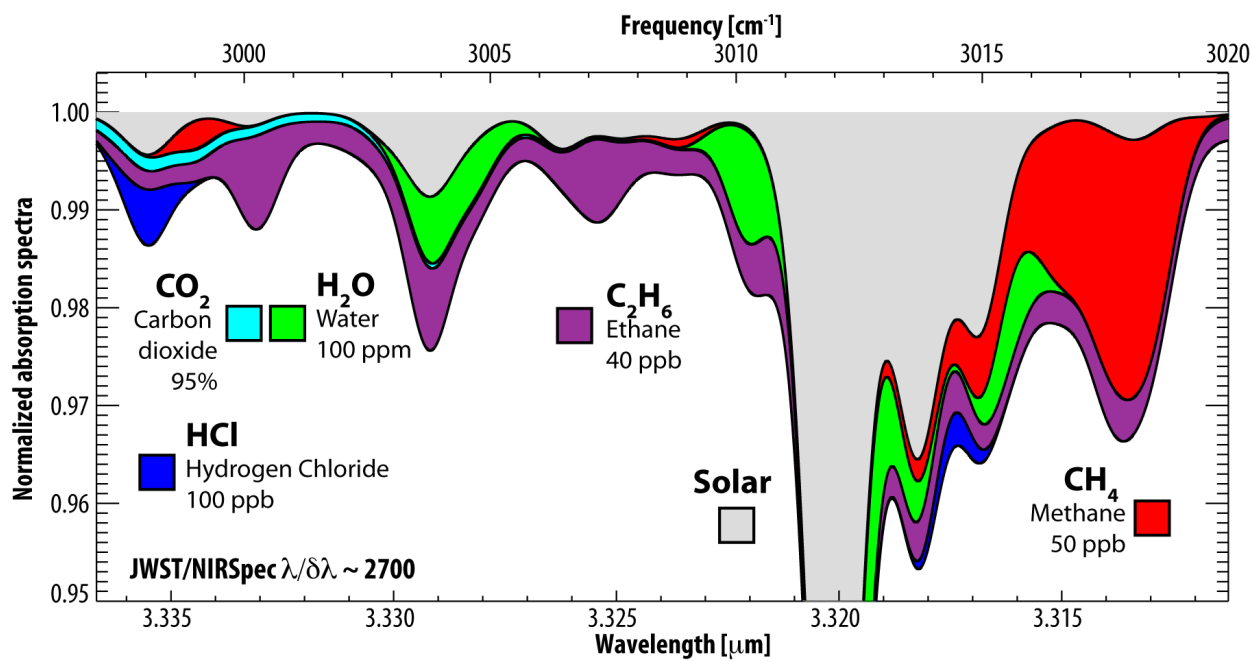

**Figure 9**: Search for organics with NIRSpec/JWST. Many trace species have strong signatures at these wavelengths (CH$_4$, C$_2$H$_6$, HCl, H$_2$O and CO$_2$ shown), enabling sensitive searches on Mars with JWST due to the observatory superb spectrometric sensitivities and high spatial resolutions.